\documentclass[letterpaper,pdftex,rmp,twocolumn,amsmath,amssymb,groupedaddress,floatfix,superscriptaddress]{revtex4}

\setcitestyle{authoryear,round}
\usepackage{natbib}

\usepackage{endnotes}
\let\footnote=\endnote
\def\footnotetext{\endnotetext[\number\numexpr\value{endnote}+1]}
\let\footnotemark\endnotemark

\usepackage[scaled]{helvet}
\usepackage[utf8]{inputenc}
\usepackage[T1]{fontenc}
\usepackage{ae}
\usepackage{aecompl}
\usepackage{color}
\usepackage{textcomp}
\usepackage{makecell}
\usepackage{mathptmx}
\usepackage{float}
\usepackage{setspace}
\usepackage[a]{esvect}
\usepackage{pdfpages}
\usepackage{placeins}

\usepackage{footnote}

\usepackage{caption}
\DeclareCaptionLabelSeparator{bar}{ \rule[-0.2em]{0.3ex}{1em} }
\usepackage[sf,bf,tiny]{titlesec}
\titlespacing*{\section}{0pt}{1em}{0em}
\titlespacing*{\subsection}{0pt}{1em}{0em}

\definecolor{darkgray}{rgb}{0.25,0.25,0.25}
\definecolor{darkred}{rgb}{0.89,0.10,0.11}
\definecolor{darkblue}{rgb}{0.12,0.39,0.62}
\usepackage{url}

\usepackage{booktabs}

\newcommand{\mytoprule}{\specialrule{0.1em}{0em}{0em}}
\newcommand{\mybottomrule}{\specialrule{0.1em}{0em}{0em}} 
\newcommand{\mymidrule}{\specialrule{0.05em}{0em}{0em}}

\usepackage{changes}

\begin{document}
	
\renewcommand{\figurename}{Figure}
\renewcommand{\thefigure}{\arabic{figure}}
\renewcommand{\tablename}{Table}
\renewcommand{\thetable}{\arabic{table}}
\renewcommand{\refname}{\large References}

\addtolength{\textheight}{1cm}
\addtolength{\textwidth}{1cm}
\addtolength{\hoffset}{-0.5cm}

\setlength{\belowcaptionskip}{1ex}
\setlength{\textfloatsep}{2ex}
\setlength{\dbltextfloatsep}{2ex}

\newcommand*{\citen}[1]{%
  \begingroup
    \romannumeral-`\x 
    \setcitestyle{numbers}%
    \citep{#1}%
  \endgroup   
}

\title{Mapping bilateral information interests using the activity of Wikipedia editors}

\author{Fariba Karimi}
\email{fariba.karimi@gesis.org}
\affiliation{Leibniz Institute for the Social Sciences, Cologne, Germany}

\author{Ludvig Bohlin}
\affiliation{Integrated Science Lab, Department of Physics, Ume{\aa} University, SE-901 87 Ume{\aa}, Sweden}

\author{Anna Samoilenko}
\affiliation{Leibniz Institute for the Social Sciences, Cologne, Germany}

\author{Martin Rosvall}
\affiliation{Integrated Science Lab, Department of Physics, Ume{\aa} University, SE-901 87 Ume{\aa}, Sweden}

\author{Andrea Lancichinetti}
\affiliation{Integrated Science Lab, Department of Physics, Ume{\aa} University, SE-901 87 Ume{\aa}, Sweden}

\begin{abstract}
\begin{minipage}{0.6\textwidth}\normalsize\sffamily

\textbf{ABSTRACT} We live in a global village where electronic communication has eliminated the geographical barriers of information exchange. The road is now open to worldwide convergence of information interests, shared values, and understanding. Nevertheless, interests still vary between countries around the world. This raises important questions about what today's world map of information interests actually looks like and what factors cause the barriers of information exchange between countries.
To quantitatively construct a world map of information interests, we devise a scalable statistical model that identifies countries with similar information interests and measures the countries' bilateral similarities. From the similarities we connect countries in a global network and find that countries can be mapped into 18 clusters with similar information interests. Through regression we find that language and religion best explain the strength of the bilateral ties and formation of clusters. Our findings provide a quantitative basis for further studies to better understand the complex interplay between shared interests and conflict on a global scale. The methodology can also be extended to track changes over time and capture important trends in global information exchange.
\end{minipage}
\end{abstract}

\maketitle

\section*{Introduction}
\noindent ``We live in a global world'' has become a clich{\'e} \citep{kose2014world}.  
Historically, the exchange of goods, money, and information was naturally limited to nearby locations, since globalization was effectively blocked by spatial, territorial, and cultural barriers \citep{cairncross2001death}. Today, new technology is overcoming these barriers and exchange can take place in an increasingly international arena \citep{friedman2000lexus}. Nevertheless, geographical proximity still seems to be important for the trade of goods \citep{overman2003economic, serrano2007patterns, fagiolo2010evolution, kaluza2010complex} as well as for mobile phone communication \citep{lambiotte2008geographical} and scientific collaboration \citep{pan2012world}. However, since the Internet allows information to travel more easily and rapidly than goods, it remains unclear what are the effective barriers of global information exchange. As information exchange requires shared interests, we therefore need to better understand global connections in interest, and the factors that form these connections.

Although globalization of information has been discussed extensively in the research literature \cite{friedman2000lexus,nye2004power,fischer2003globalization}, currently there is no method to quantitatively map bilateral information interests from large-scale data.
Without such a method, it becomes difficult to justify qualitative statements about, for example, the complex interplay between shared values and conflict on a global scale. We use data mining and statistical analysis to device a measure of bilateral information interests, and use this measure to construct a world map of information interests.




To study interests on a global scale, we use the free online encyclopedia Wikipedia, which has evolved into one of the largest collaborative repositories of information in the history of mankind \cite{mesgari2014sum}. The free online encyclopedia consists of almost 300 language editions, with English being the largest one\footnote{\url{http://en.wikipedia.org/wiki/Wikipedia:Size_of_Wikipedia}\footnotemark}. This multi-lingual encyclopedia captures a wide spectrum of information in millions of articles. These articles undergo a peer-reviewed editing process without a central editing authority. Instead, articles are written, reviewed, and edited by the public. Each article edit is recorded, along with a time-stamp, and, if the editor is unregistered, the computer's IP address. The IP address makes it possible to connect each edit to a specific location. Therefore we can use Wikipedia editors as sensors for mapping information interest to specific countries. 

In this paper, we use co-editing of the same Wikipedia article as a proxy for shared information interests. To find global connections, we look at how often editors from different countries co-edit the same articles. To infer connections of shared interest between countries, we develop a statistical model and represent significant correlations between countries as links in a global network. Structural analysis of the network suggests that interests are polarized by factors related to geographical proximity, language, religion and historical background. We quantify the effects of these factors using regression analysis and find that information exchange indeed is constrained by the impact of social and economic factors connected to shared interests.

\section*{Methodology}
\section*{Relating information interests to geographical location}
As one of the largest and most linguistically diverse repositories of human knowledge, Wikipedia has become the world's main platform for archiving factual information \cite{mesgari2014sum}. One important feature of Wikipedia is that every edit made to an article is recorded.  Thanks to this detailed data, Wikipedia provides a unique platform for studying different aspects of information processes, for example, semantic relatedness of topics \cite{radinsky2011word, auer2007have}, collaboration \cite{kimmons2011understanding,torok2013opinions,keegan2012editors}, social roles of editors \cite{welser2011finding}, and the geographical locations of Wikipedia editors \cite{lieberman2009you}. 

In this work, we used data from Wikipedia dumps\footnote{Available on \url{http://dumps.wikimedia.org/enwiki/}\footnotemark} to select a random sample from the English Wikipedia edition, which is the largest and most widespread language edition. In total, the English edition has around 10 million articles, including redirects and duplicates. Since retrieving the editing histories of all articles is computationally demanding, we randomly sampled more than six million articles from this set. For each English article, we retrieved the complete editing history of the same article in all language editions that the English Wikipedia page links to. Finally we merged all language editions together to create a global editing history for each article. For each edit, the editing history includes the text of the edit, its time-stamp, and, for unregistered editors, the IP address of the editor's computer. From the IP address associated with the edit, we retrieved the geolocation of the corresponding editor using an IP database \footnote{We used \url{http://www.ip2location.com/}\footnotemark}. For the purpose of spatial analysis, we limited the analysis to edits from unregistered editors, because data on the location for most of the registered Wikipedia editors are unavailable. The resulting dataset contains more than six million (6,285,753) Wikipedia articles and about 140 million edits in total. We use these edits to create interest profiles for countries.



\section*{Inferring shared interests from edit co-occurrence} 

We identify the interest profile of a country by aggregating the edits of all Wikipedia editors whose IPs are recorded in the country. If an article is co-edited by editors located in different countries, we say that the countries share a common interest in the information of the article. In other words, we connect countries if their editors co-edit the same articles. Indirectly, we let individuals who edit Wikipedia represent the population of their country. While Wikipedia editors in a country certainly do not represent a statistically unbiased sample, there is a higher tendency that they edit contents that are related to the country in which they live \cite{hecht2010}. Therefore, we approximate the interest profile of a country with collective editing behavior of editors in that country. 



Inferring the location of all editors on the country level is non-trivial. Although we have data on all edits, we do not know the location of registered editors because their IPs are not recorded. One proposed approach to tackle this problem makes use of circadian rhythms of editing activity to infer the location of the editors \cite{yasseri2012circadian}. This method approximates the longitude of a location but provides little information about its latitude. Therefore, we must limit the analysis to the activity of unregistered editors with recorded IP addresses. This will arguably affect the results. Not only do registered editors contribute to 70\% of all 140 million edits, they also have somewhat different behavior. For example, many of the most active registered users take on administrative functions, develop career paths, or specialize in covering selected topics \cite{arazy2015functional}. On the other hand, some unregistered editors are involved in vandalism, but often their activity nevertheless indicates their interest.\footnote{See Wikipedia's policy and fight against vandalism here: \url{https://en.wikipedia.org/wiki/Vandalism_on_Wikipedia}} While we can only speculate about how including registered editors would affect the results, unregistered editors can nevertheless provide useful information about shared interests between countries.

From the co-editing data, we create a network that represents countries as nodes and shared interests as links. The naive approach is to use the raw counts of co-edits between countries as weighted links. The problem with this approach is that it is biased toward the number of editors in each country. Some countries may be strongly connected, not because of evident shared interests but merely as a result of a large community of active Wikipedia editors. To address this problem, we propose a statistical validation method that filters out connections that could exist only due to size effects or noise. The filtering method assumes a multinomial distribution and determines the expected number of co-occurring edits from the empirical data. In other words, we infer significant links in a bipartite system in which countries are in one set and articles are in the other set. There are other existing methods to evaluate the significant correlation between entities in bipartite systems. For example, \citet{zweig2011systematic} proposed a systematic approach to one-mode projections of bipartite graphs for different motifs. In another work, \citet{tumminello2011statistically} used the hypergeometric distribution and measured the $p$-value for each subset of the bipartite network. Moreover, \citet{lancichinetti2014high} proposed a community detection method to classify topics to articles more efficiently, and \citet{serrano2009extracting} used a disparity filtering method to infer significant weights in networks. Finally, \citet{Ronen15122014} adopted a statistical approach to determine significant links between languages in various written documents. However, the model that we use has the advantages that it makes it easy to account for size variation inside an article and to compute the $z$-scores for analyzing the country-based editor activity.

\section*{Interest model} 
\begin{figure}[tbp]
\begin{center}
\includegraphics[width = 0.5 \textwidth]{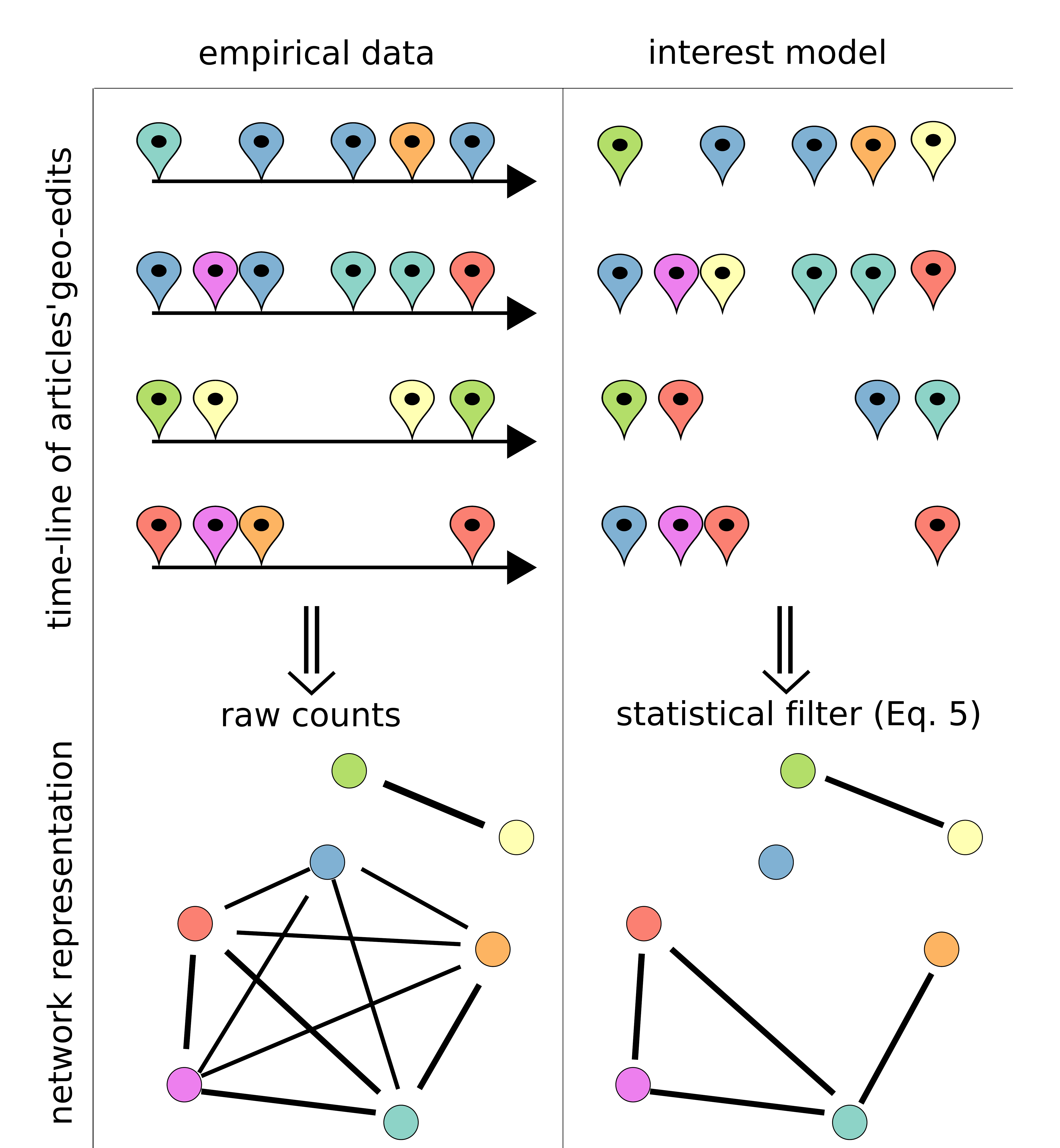}
\caption{Illustration of the interest model. The left panel shows the time-ordered edit sequence of four Wikipedia articles, with edits coming from different countries represented as colored pins. Note that pins represent country edits, and therefore they can be repeated. The resulting empirical network, calculated by multiplying raw co-edits counts, is at the bottom. In the right panel, we illustrate the null model with the same four articles, and the resulting network at the bottom. In the null model, the average editing activity of the countries is preserved, but the order of the edits is reshuffled within and across articles. As a result of the filtering, some links are removed in the interest model. \label{fig:model}}
\end{center}
\end{figure}
In this section we outline the formalisation of the model. We link countries based on their co-occurring edits over all Wikipedia articles. For a specific article $a$, we calculate the link weight between all pairs of countries that edited the article, as follows: if editors in country $i$ have edited an article $k_{i}^{a}$ times, and editors in country $j$ have edited the same article $k_{j}^{a}$ times, then the countries' empirical link weight, $w_{ij}^{a}$, is calculated as:

\begin{equation}
w_{ij}^{a} = k_{i}^{a} k_{j}^{a}.
\label{eq:M1_empirical}
\end{equation}
Since the total number of articles is over six million, most country pairs have co-edited at least one article. Therefore, the aggregation of all articles results in numerous links between countries, and the countries with relatively large editing activities become highly central. 
Accordingly, we cannot know if the link exists by chance, or because countries actually tend to edit the same articles more frequently than expected. To determine which links are statistically significant, we compare the empirically observed link weights with the weight given by a null model. In the null model, we assume that each edit comes from a country randomly picked proportionally to its total number of edits. More specifically, the random assignments are performed by drawing the countries from a multinomial distribution. That is, for each edit, country $i$ is selected proportional to its cumulative editing activity, $p_{i} = \frac{\sum_{a}k_{i}^a}{M}$, where $M$ is the total number of edits for all articles. Note that each edit is sampled independently from all other edits, and that the cumulative edit activity of a country in the null model on average will be the same as the observed one. This null model preserves the average level of activity of the countries, but randomizes the temporal order and the articles that countries edit. Figure~\ref{fig:model} shows an example of this reshuffling scheme with four articles. 

From the null model, we can analytically compute the expected probability, $\mu_{ij}^{a}$, that two countries $i$ and $j$ edit the same article $a$ (detailed derivation in the Methods):
\begin{equation}
\mu_{ij}^{a} = n_{a} (n_{a}-1) p_{i} p_{j},
\label{eq:M1_1}
\end{equation}
where $n_{a}$ is the total number of edits in article $a$.

To compare the empirical and expected link values, we compute standardized values, so called $z$-scores. For countries $i$ and $j$ and article $a$, the $z$-score $z_{ij}^{a}$ is defined as
\begin{equation}
z_{ij}^{a} = \frac{w_{ij}^{a} - \mu_{ij}^{a}}{\sigma_{ij}^{a}},
\label{eq:M1_2a}
\end{equation}
where the standard deviation $\sigma_{ij}^{a}$, is computed in the Method section.

The $z$-scores are useful for comparisons of weights, since they account for the large variations that exist in the articles' edit histories. We then sum over all articles to find the cumulative $z$-score for countries $i$ and $j$ 
\begin{equation}
z_{ij} = \sum_{a} \frac{w_{ij}^{a} - \mu_{ij}^{a}}{\sigma_{ij}^{a}}. 
\label{eq:M1_2}
\end{equation}


Using the Bonferroni correction, we consider a link to be significant if the probability of observing the total $z$-score is less than $0.05 / N$, where $N$ is the number of countries. Since the total $z$-score is a sum over many independent variables, we can approximate the expected total $z$-score distribution with a normal distribution. The normal distribution has average value $0$ and standard deviation $\sqrt{L}$, where $L$ is the number of Wikipedia articles. Thus, the threshold for the significant link weight is $t = 3.52 \sqrt{L}$, where $3.52$ is derived from the condition that $P(z > 3.52) = 0.05 / N$, where $N=234$ countries and $P$ is the standard Gaussian distribution (with zero average and unit variance). If the total $z$-score is larger than the threshold, we create a link between countries $i$ and $j$ with weight $\widetilde{w_{ij}}$ according to
\begin{equation}
\widetilde{w_{ij}} =
  \begin{cases}
   z_{ij}-t & \text{if } z_{ij} > t \\
   0 & \text{if } z_{ij} \leq t.
  \end{cases}
\end{equation}

In summary, the interest model maintains the average level of activity of the countries and randomizes the articles that they edit. By comparing results from the interests model and empirical values, we can identify significant links between countries.


\begin{figure*}
\begin{center}
\includegraphics[width=0.9\textwidth]{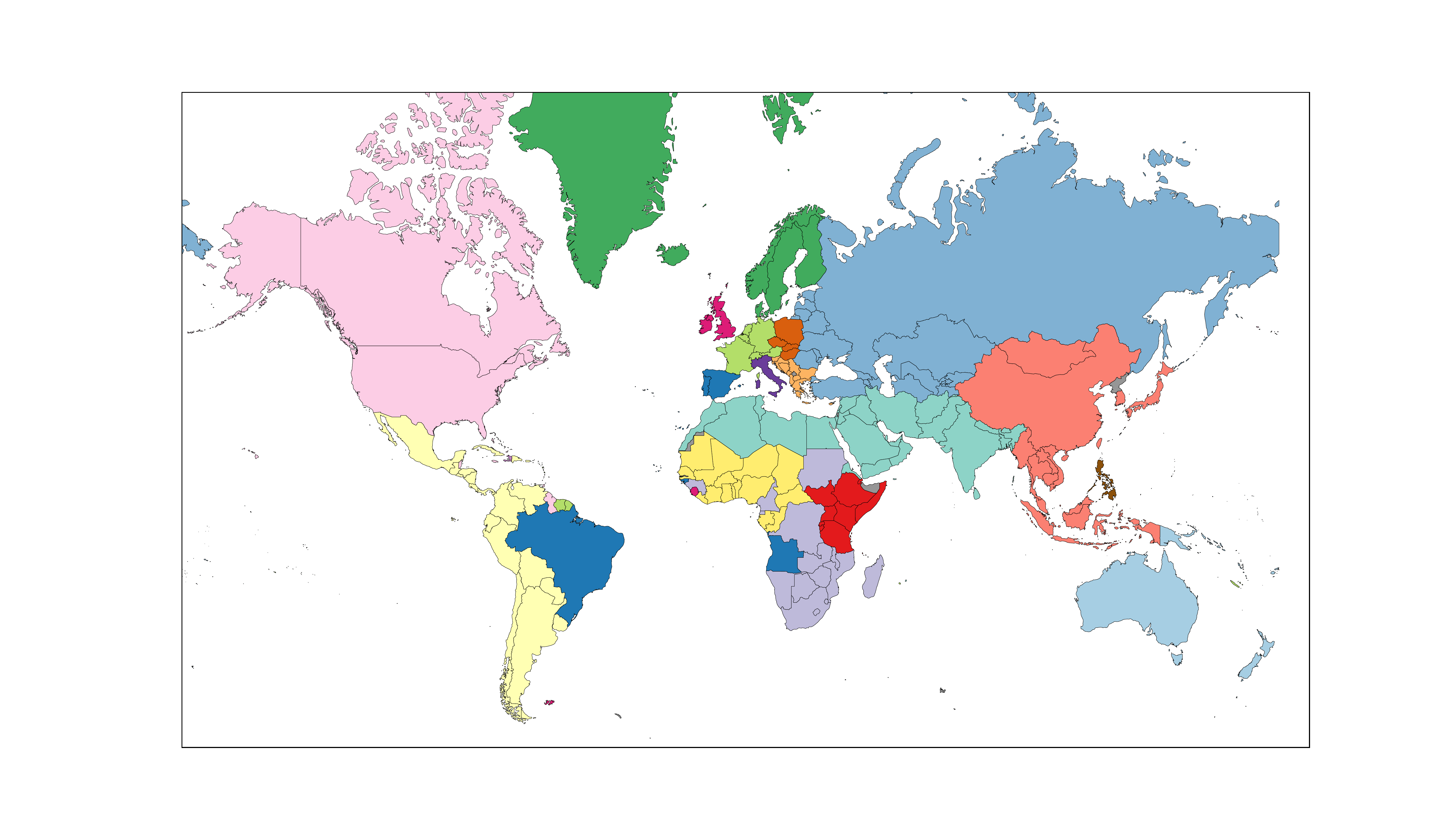}

\caption{World map of information interests. Countries that belong to the same cluster have the same color. Countries colored with gray do not belong to any cluster. The map suggests that the division of countries can be explained by a combination of cultural and geopolitical features. \label{fig:full_worldmap_M1}}
\end{center}
\end{figure*}

\begin{figure*}
\begin{center}
\includegraphics[width=0.75\textwidth]{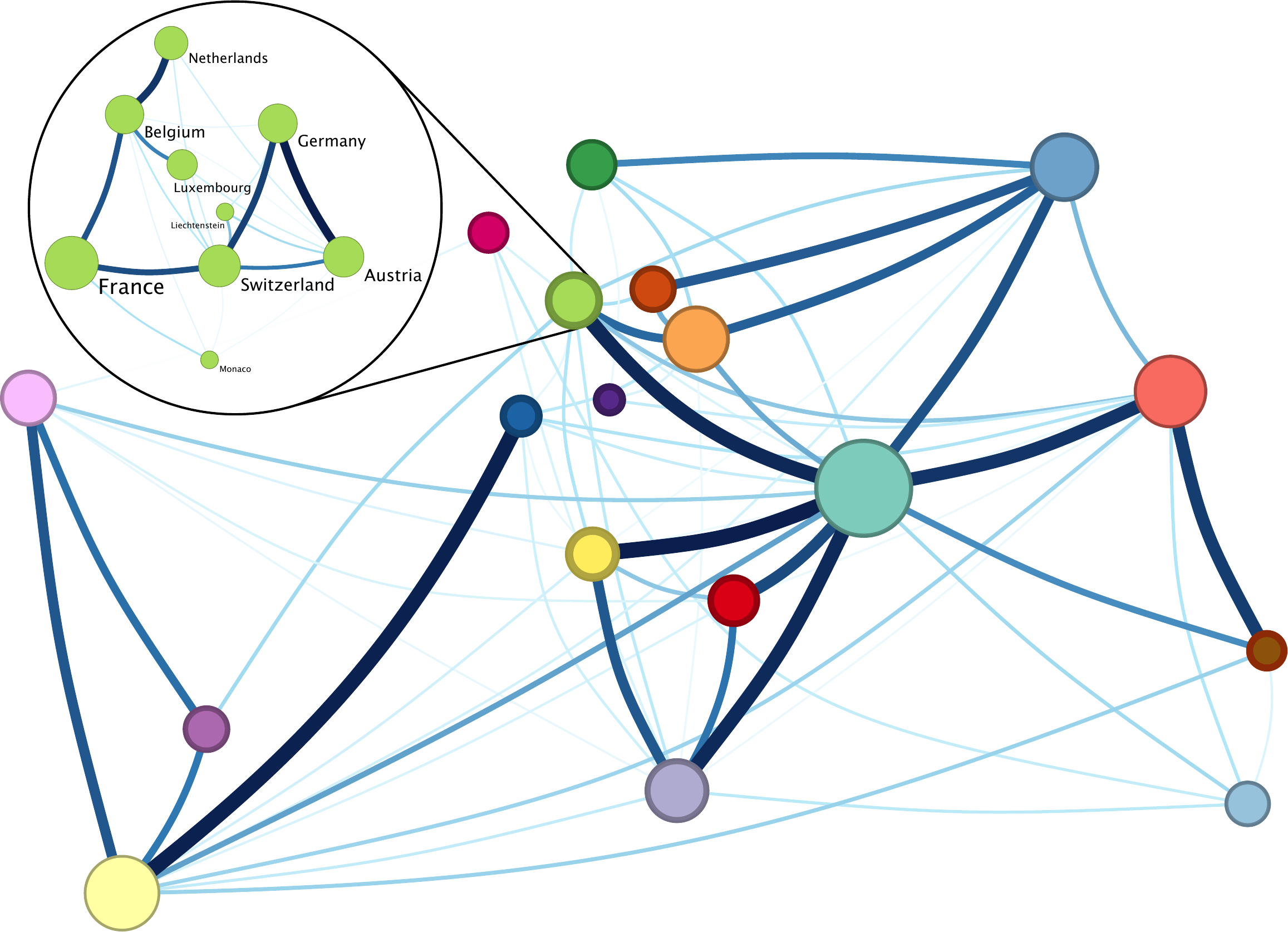}

\caption{World network of information interests.
The size of the nodes represents the total $z$-score of the clusters. The links represent connections between clusters obtained from the cluster analysis with Infomap, the thicker the line, the stronger the connection. Clusters are colored in the same way as in Fig.~\ref{fig:full_worldmap_M1}. The upper left corner shows the most significant connections between countries in the Central European cluster.
\label{fig:worldfrance} }
\end{center}
\end{figure*}

\section*{Clustering countries with similar interests}
To investigate the effective barriers of global information exchange, we first identify large-scale structures among the thousands of links between countries. In this way, we can highlight the groups of countries that share interest in the same information. To reveal such groups among the pairwise connections, we use a network community detection method based on random walks as a proxy for interest flows. While the community-detection method we use is good at breaking chains of links, we may connect some countries primarily based on strong connections with common countries and not between themselves. Nevertheless, simplifying and highlighting important structures provide a valuable map to investigate the large-scale structure of global information exchange.

In our clustering approach, we first build a network of countries connected with the significant links we find in our filtering. To identify groups of countries, we envision an editor game in which editors from different countries are active in sequence. In this relay race, a country exchanges information to another country proportional to the weight of the link between the countries. Accordingly, the sequence of countries forms a random walk and certain sets of countries with strong internal connections will be visited for a relatively long time.  This process is analog to the community-detection method known as the map equation \cite{rosvall2008maps, rosvall2009map}. Here we use the map equation's associated search algorithm Infomap \cite{mapequation} to identify the groups of countries we are looking for and to reveal the large-scale structure.

\section*{Results and discussion}
We will discuss the results at four levels of detail, from the big picture to the detailed dynamics, and highlight different potential mechanisms for barriers of information exchange. First, we will show a global map of countries with shared information interests, and continue with the interconnections between the clusters. Then we will consider each cluster separately and examine the interconnections between countries within the clusters. Finally, we will apply multiple regression analysis to examine explanatory variables to the highways and barriers of information exchange.

\subsection*{The world map of information interests}

The world map of information interests suggests that cultural and geopolitical features can explain the division of countries.
Between the 234 countries, we identified 2,847 significant links that together form a network of article co-edits. By clustering the network, we identified 18 clusters of strongly connected countries (see Supplementary Table for a detailed list of countries in each cluster). The resulting network is illustrated as a map in Fig.~\ref{fig:full_worldmap_M1}, where countries of the same cluster share the same color. The map suggests that the division of countries can be explained by a combination of cultural and geopolitical features. For example, the United States and Canada share a long geographical border and extensive mutual trade, and are clustered together despite the fact that other English-speaking countries are not. Moreover, religion is a plausible driver for the formation of the cluster of countries in the Middle East and North Africa, as well as the cluster of Russia and the Orthodox Eastern-European countries \cite{gupta2002cultural}. Another factor in the formation of shared information interests is language. For example, countries in Central and South America are divided into two clusters with Portuguese and Spanish as common languages in each cluster, respectively. Colonial history can also shape similarity in interests, as in the cluster of Portugal, Angola and Brazil, as well as the cluster of former Soviet Union countries \cite{colonial}. Overall, there is strong empirical evidence that geographical proximity, common religion, shared language, and colonial history can explain the division of countries.

To examine the connections between clusters, we looked at the network structure at the cluster level. The network in Fig.\ref{fig:worldfrance} shows the connections between the clusters of countries illustrated in Fig.~\ref{fig:full_worldmap_M1} with the same color coding. Connections tend to be stronger between clusters of geographically proximate countries also at this level. Interestingly, the Middle East cluster in turquoise has the largest link strengths to other clusters, forming a hub that connects East and West, North and South. Interpreting the strong connections as potential highways for information exchange, the Middle East is not only a melting pot of ideas, but also plays an important role in the spread of information.



To get better insights into how the clusters are shaped, we zoomed into the country networks within clusters. In the upper left corner of Fig.~\ref{fig:worldfrance}, we show the strongest connections within the Central European cluster. It suggests that countries are linked based on the overlap of the official languages \cite{hale2014multilinguals}. For example, Belgium has three official languages, Dutch, French and German. Indeed, Belgium is connected closely with the Netherlands, France, and Luxembourg. We observed the same pattern in other clusters, and the triad of Switzerland, Germany, and Austria is another example of strongly linked countries with a shared language.
 

Just to illustrate what interests can form the bilateral connections, we looked at a number of concrete examples. First, we ranked the articles according to their significant $z$-scores for each pair of countries. Then we looked at the top-ranked articles and here report the results for two European country pairs: Germany--Austria in the European cluster and Sweden--Norway in the Scandinavian cluster. The articles with the most significant co-edits relate to local and regional interests, including sports, media, music, and places (see Supplementary Table). For example, the top-ranked articles in the Germany--Austria list include an Austrian singer who is also popular in Germany, and an Austrian football player who is playing in the German league. The top-ranked articles in the Sweden--Norway list shows a similar pattern of locally related topics, for example, a host of a popular TV show simultaneously aired in Sweden and Norway, a Swedish football manager who has been successful both in Sweden and Norway, and a music genre that is nearly exclusive to Scandinavian countries. Altogether, the top articles suggest that an important factor for co-editing is related interests, which in turn may be an effect of shared language, religion, or colonial history, as well as geographical proximity or large volume of trade between countries.



\subsection*{Regression analysis of the highways and barriers of information exchange}

To quantify the impact of social and economic factors behind the shared interests, we performed a Multiple Regression Quadratic Assignment Procedure (MRQAP) analysis. This method is specifically suited when there are collinearity and autocorrelation in the data \cite{krackhardt1988predicting, dekker2007sensitivity}. We performed the MRQAP using the \texttt{netlm} function in the \texttt{sna} R package \cite{butts2008social}. The dependent variables in the regression model were the significant $z$-scores that we obtained from the data. Our independent variables were geographical proximity, trade \cite{subramanian2007wto}, colonial ties\footnote{We used ICOW Colonial History Data Set, version 1.0. available on \url{http://www.paulhensel.org/icowcol.html}}, shared language\footnote{We used Ethnologue available on \url{http://www.ethnologue.com/}}, and shared religion\footnote{We used the World Religion Database available on \url{http://www.worldreligiondatabase.org/}}, as suggested by the analysis of the map of information interests (see the Supplementary Information S2 for a more detailed description of the data).

\begin{table*}[htb]
\caption{\label{table:regression}\textbf{Results of the multiple regression analysis.} Significant edit co-occurrences ($z$-scores) form the dependent variable matrix, which we regress on the independent matrices in different models. Values in parenthesis are $t$-statistics. The features are ordered by importance, from shared language to trade. Country pairs = 62,001. Values marked with an asterisk have a $p$-value less than 0.01}
\centering
\setlength{\tabcolsep}{2.9pt}
\begin{tabular}{lrrrrr}
\mytoprule\noalign{\smallskip}
                   & $R_{0}$    & $R_{1}$    & $R_{2}$      & $R_{3}$      & $R_{4}$      \\ \mymidrule\noalign{\smallskip}
                   
Intercept          & 0.41       & 0.3        & 2.33         & 2.33         & 2.28         \\ 
Shared language    & 0.91$^*$ (69) & 0.82$^*$ (64) & 0.77$^*$ (60)   & 0.75$^*$ (58)   & 0.74$^*$ (57)   \\ 
Shared religion    &            & 2.76$^*$ (46) & 2.6$^*$ (44)    & 2.6$^*$ (43)    & 2.44$^*$ (40)   \\ 
Log distance       &            &            & -0.23$^*$ (-23) & -0.23$^*$ (-23) & -0.23$^*$ (-23) \\ 
Colonial tie       &            &            &              & 4.5$^*$ (22)    & 4.35$^*$ (21)   \\ 
Log trade          &            &            &              &              & 0.03$^*$ (10)    \\ \noalign{\bigskip}
Adjusted R-squared & 0.13       & 0.19       & 0.20         & 0.21         & 0.22         \\ 
F-statistic        & 7,774       & 3,590       & 2,610         & 2,110         & 1,716         \\ 
dF                 & 30,874      & 30,873      & 30,872        & 30,871        & 30,870        \\ \mybottomrule
\end{tabular}
\end{table*}

All independent variables show significant correlation with the data (see Table \ref{table:regression}). To observe the variation between different independent matrices, we examined them in different models. In model $R_{0}$, we examined
the influence of shared language, which explains 13\% of our observation. In model $R_{1}$, we added shared religion, which increases the power of the model to 19\%. In model $R_{2}$, we included the geographical proximity. It slightly increases the R-squared and has negative relation to the observed $z$-scores, since short distance corresponds to high proximity. In models $R_{3}$ and $R_{4}$, respectively, we added colonial ties and trade. Including all these explanatory variables into the regression model enabled us to increase the explanatory power of the model to 22\%. The correlation of each variable with the observed $z$-scores can be inferred from the $t$-statistic. Shared language shows the strongest association, followed by shared religion, geographical proximity, colonial ties, and volume of trade (see Table \ref{table:regression}).

The influence of language on shared interests is not surprising. It is well known that interests are formed by cultural expression and public opinion, and language is an important platform for these expressions \cite{usunier2005marketing}. That the relation between language and interests is important has also been demonstrated by the surprisingly small overlap between languages in Wikipedia \cite{hecht2010tower, callahan2011cultural} and the variation in the editing of controversial topics \cite{yasseri2014most}. 

Moreover, the influence of religion is in line with the Huntington's thesis that the source of division between people in the post-Cold War period is primarily rooted in cultural differences and religion \cite{huntington1993clash}. Similar results were found in other studies that analyzed Twitter and email communication worldwide \cite{park2015mesh}.

Overall, the analysis reveals that information exchange is constrained by the impact of social and economic factors connected to shared interests. In other words, globalization of the technology does not bring globalization of the information and interests. Language, religion, geographical proximity, historic background, and trade are potential driving factors to polarize the information interests. These results coincide with earlier works that highlight the impact of the colonization, immigration, economics, and politics on the cultural similarities and diversities \cite{feldman2001brazilians,bleich2005legacies,gelfand2011differences,castells2011power,tagil1995ethnicity,risse2001european,hennemann2012myth}.


%

%
%

\section*{Conclusions}
By simplifying and highlighting the important structure in the myriad edits of Wikipedia, we provide a world map of shared information interests. We find that information interests despite globalization are diverse, and that the highways and barriers of information exchange are formed by social and economic factors connected to shared interests. In descending order, we find that language, religion, geographical proximity, historic background, and trade explain the diversity of interests. While technological advances in principle have made it possible to communicate with anyone in the world, these social and economic factors limit us from doing so and information interests remain diverse. Questions remain how different social and economic factors affect different regions, how they relate to conflicts on a global scale, and how the impact of these factors changes over time. It would therefore be interesting to extend the methodology to track changes over time.

\section*{Method}
To find connections in interest, we measure the co-occurring edits of two countries in the same articles. We quantify the connection with an empirical weight that is computed as the product of the countries' edit activities in the article. For a Wikipedia article $a$, if the total edit activity of country $i$ is denoted by $k_{i}^{a}$, and for country $j$ is $k_{j}^{a}$, then we calculate the empirical weight, $w_{ij}^{a}$, according to
\begin{equation}
w_{ij}^{a} =  k_{i}^{a} k_{j}^{a}.
\end{equation}
As the total edit activity of countries differs, the probability that countries appear together in a certain article varies. If the total number of edits for all articles is $M$, then the expected proportion of edits for country $i$ is 
\begin{equation}
p_{i} = \frac{\sum_{a}k_{i}^a}{M}.
\end{equation}
This is the probability of country $i$  making a random edit overall, and this is the null model we use to filter noisy connections in the interest model. Assume that there is a total of $c$ countries, and a total of $n$ edits for article $p$. Let $x_{k}$ denote the number of edits from country $k$. Then the probability of any particular combination of edits for the various countries follows a multinomial distribution
\begin{equation}
 \frac{n!}{x_{1}!... x_{c}!} p_{1}^{x_{1}} ... p_{c}^{x_{c}},  \textrm{where} \sum\limits_{i=1}^c x_{i} = n.
\end{equation}
With the above distribution, we can compute the expected probability of the co-occurrence of two countries $i$ and $j$ in an article
\begin{equation}
\mu_{ij}^{a} = \sum_{1...k } x_{i} x_{j} \frac{n!}{x_{1}!... x_{n}!} p_{1}^{x_{1}} ... p_{n}^{x_{n}}
\end{equation}
Following the multinomial theorem, we can now calculate the mean, variance and covariance matrices for the occurrence of a country pair. The mean value of the co-occurrence of two countries becomes
\begin{equation}
\mu_{ij}^{a} = n_{a} (n_{a}-1) p_{i} p_{j}.
\end{equation}
Using the multinomial theorem multiple times, one can also compute the variance:
\begin{align}
\label{eq:variance}
(\sigma_{ij}^{a})^2 =& n_{a} (n_{a}-1) p_{i} p_{j} ( (6-4n_{a}) p_{i} p_{j} + (n_{a}-2) (p_{i}+p_{j}) + 1 ).
\end{align}
Thus, the standard deviation of the pair, $\sigma_{ij}^{a}$, is the square root of equation \eqref{eq:variance}.
This equation enters into the definition of the $z$-score in equation \eqref{eq:M1_2}. The sum of the $z$-scores is then approximated with a Gaussian distribution. This approximation is well justified by the large number of articles. In practice, already $1,000$ articles give a good approximation, as shown by the numerical simulations in the Supplementary Information.

\bibliographystyle{apalike}


\section*{\large Acknowledgements} We thank Alcides V.~Esquivel,  Daniel Edler, Claudia Wagner, Markus Strohmaier and Micheal Macy  for valuable discussions. We also thank the Wikimedia Foundation for providing free access to the data. F.K., L.B., A.L. and M.R.\ were supported by the Swedish Research Council grant 2012-3729.
%
%
\section*{\large Additional information} The authors declare that they have no competing financial interests. Correspondence and requests for materials should be addressed to F.K.~\mbox{(fariba.karimi@gesis.org)}. The datasets generated during and/or analysed during the current study are available in the ``google sites'' repository, `` https://sites.google.com/site/mappingbilateralwiki/''.

\FloatBarrier
\clearpage



\section*{Supplementary material}

**Supplementary Table 1 | Clustering results.** In total 234 countries assigned to 18 clusters

| Cluster | Countries |
|---|---|
| 1 | Saudi Arabia, United Arab Emirates, Egypt, India, Kuwait, Jordan, Qatar, Pakistan, Bahrain, Palestine, Oman, Algeria, Morocco, Lebanon, Syria, Iraq, Tunisia, Yemen, Bangladesh, Libya, Sri Lanka, Iran, Nepal, Maldives, Israel, Mauritius, Afghanistan, Bhutan, Eritrea |
| 2 | Argentina, Colombia, Venezuela, Guatemala, Peru, Mexico, Chile, Ecuador, Uruguay, Honduras, Costa Rica, Dominican Republic, Panama, Paraguay, El Salvador, Bolivia, Puerto Rico, Nicaragua, Cuba |
| 3 | Hong Kong, Taiwan, China, Malaysia, Singapore, South Korea, Vietnam, Indonesia, Thailand, Macau, Japan, Cambodia, Burma (Myanmar), Brunei, Mongolia, Laos, Timor-Leste |
| 4 | Russian Federation, Ukraine, Belarus, Azerbaijan, Kazakhstan, Latvia, Armenia, Estonia, Georgia, Lithuania, Moldova, Romania, Turkey, Uzbekistan, Kyrgyzstan, Turkmenistan, Tajikistan |
| 5 | Serbia, Bosnia and Herzegovina, Montenegro, Croatia, Macedonia, Greece, Bulgaria, Slovenia, Cyprus, Albania |
| 6 | South Africa, Sudan, Zimbabwe, Cameroon, Democratic Republic of the Congo, Botswana, Zambia, Namibia, Mozambique, Swaziland, Equatorial Guinea, Guinea, Madagascar, Malawi, Lesotho, Sao Tome and Principe |
| 7 | France, Switzerland, Austria, Germany, Belgium, Netherlands, Luxembourg, Monaco, Suriname, Liechtenstein, French Polynesia, New Caledonia, Mayotte, Réunion, Saint Pierre and Miquelon |
| 8 | United States, Canada, Bermuda, Palau, Bahamas, Caribbean Islands[1] |
| 9 | Nigeria, Senegal, Ghana, Ivory Coast, Burkina Faso, Benin, Mauritania, Mali, Liberia, Niger, Gambia, Gabon, Togo, Republic of the Congo, Chad, Central African Republic |
| 10 | Kenya, Uganda, Djibouti, Somalia, Tanzania, Rwanda, Ethiopia, South Sudan, Burundi, Comoros |
| 11 | Sweden, Denmark, Norway, Finland, Greenland, Faroe Islands, Iceland, Åland Islands, Malta |
| 12 | Curaçao, Saint Martin, Guadeloupe, Sant Maarten, French Guiana, Aruba, Martinique, Haiti, Wallis and Futuna |
| 13 | Slovakia, Czech Republic, Hungary, Poland, Niue |
| 14 | Fiji, New Zealand, Australia, Samoa, Vanuatu, Kiribati, Cook Islands, Tonga, Solomon Islands, Papua New Guinea, Nauru, Marshall Islands, American Samoa, Norfolk Island |
| 15 | Spain, Portugal, Angola, Brazil, Cape Verde, Andorra, Guinea-Bissau |
| 16 | United Kingdom, Ireland, Guernsey, Jersey, Isle of Man, Sierra Leone, Gibraltar, Falkland Islands, Tuvalu, British Indian Ocean Territory |
| 17 | Philippines, Guam, Northern Mariana Islands, Micronesia |
| 18 | Italy, San Marino, Holy See (Vatican City) |

**Supplementary Table 2 | Top 10 Wikipedia articles that Germany-Austria and Sweden-Norway co-edit based on the filtering analysis**

| Rank | DE-AT | SE-NO |
|---|---|---|
| 1 | Christina Stürmer | Tipuloidea |
| 2 | Erste Allgemeine Verunsicherung | Dansband |
| 3 | Steffen Hofmann | Boyoz |
| 4 | Nazar (rapper) | Fredrik Skavlan |
| 5 | Piefke | Erik Hamrén |
| 6 | ATV (Austria) | Sweden |
| 7 | Klagenfurt | Anders |
| 8 | Karl-Heinz Grasser | Peter Jöback |
| 9 | Wolf Haas | Causerie |
| 10 | Zillertal | List of the busiest airports in the Nordic countries |

## S2 Data for the regression analysis

**Geographical proximity** The geographical proximity is calculated as the euclidean distance between each pair of countries using Haversine formula. The point for each country is based on the rounded latitude and longitude of the centroid or the center point of the country[1].

**Trade data** Data on free trade areas and customs union (FTAs) are collected for the year 2000[2]. WTO members are obliged to notify the regional trade agreements in which they participate. 157 countries reported their trade flow in 2000. The trade volume is estimated by averaging the import and export volume of each pairs of countries.[3].

**Colonial ties** We used Colonial history dataset available to download at[4]. We create a tie between two countries if they had a colonial relationship since 1900 until now.

**Language similarities** The data from languages that are spoken in each countries are collected from Ethnologue[5]. The data contains information on more than 7000 known living languages in the world. It is regarded to be the most comprehensive source of information about languages. From the data, we weight each pair of countries dependent on number of languages that are co-spoken.

**Shared religion** The data on the religion composition of countries was taken from World Religion database[6]. The results is a binary matrix where countries that share the same religion have non-zero entities.



\end{document}